\documentclass{amsart}
\usepackage{amssymb}     
\def\R{\mathbb{R}}

\def\epsi{\varepsilon}
\newtheorem{The}{Theorem}[section]

\newtheorem{Cor}[The]{Corollary}
\newtheorem{Lem}[The]{Lemma}
\theoremstyle{plain}
\newtheorem*{Con}{Condition}\newtheorem*{Rem}{Remark}
\numberwithin{equation}{section}
\begin{document}
\title[Semiclassical Limit with a Short Scale Periodic Potential]{Semiclassical Limit for the 
Schr\"{o}dinger Equation with a Short Scale Periodic Potential}
\author{Frank H\"{o}vermann, Herbert Spohn, Stefan Teufel}
\address{Zentrum Mathematik\\ Technische Universit\"{a}t M\"{u}n\-chen\\ D-80290 
M\"{u}n\-chen\\ Germany}
\email{spohn@ma.tum.de, teufel@ma.tum.de}
\begin{abstract}
We consider the dynamics generated by the Schr\"{o}ding\-er operator 
$H=-\frac{1}{2}\Delta+V(x)+W(\epsi x)$, where $V$ is a lattice periodic potential 
and $W$ an external potential which varies slowly on the scale set by the lattice spacing. 
We prove that in the limit $\epsi\to 0$ the time dependent position operator 
and, more generally, semiclassical observables
converge strongly to a limit which is determined by the semiclassical dynamics.
\end{abstract}
\maketitle
\section{Introduction}
A basic problem of solid state physics is to understand the motion of electrons in the 
periodic potential which is generated by the ionic cores. 
While this problem is quantum mechanical, 
many electronic properties of solids can be understood already in the semiclassical 
approximation \cite{ashcroft-mermin,kohn,zak}. One argues that 
if the wave packet spreads over many lattice spacings,
the kinetic energy $(\hbar k)^2/2m$ is modified to the $n$-th band energy $E_n(k)$. Otherwise 
the electron responds to external fields, $E_\mathrm{ex}$, $B_\mathrm{ex}$,   
as in the case of vanishing periodic potential. Thus the semiclassical equations of motion are
\begin{equation} \label{semiclassical_dynamics}
\begin{array}{l} \displaystyle{
\dot{r}=v_n(k)=\nabla_k E_n(k)}\\ \\\displaystyle{
\hbar\dot{k}=e(E_{\mathrm{ex}}(r)+v_n(k)\wedge B_{\mathrm{ex}}(r))\,,}
\end{array}
\end{equation}
where $r$ is the position and $k$ the quasimomentum of the electron. Note that there is 
a semiclassical evolution for each band separately.

The goal of our paper is to understand on a mathematical level how these semiclassical 
equations arise from the underlying Schr\"{o}dinger equation. 
We consider only the case where $B_{\mathrm{ex}}=0$.

The setup is rather obvious. We start from the Schr\"{o}dinger equation
\begin{equation}\label{basic_dynamics}
i\frac{\partial}{\partial t}\psi=H\psi
\end{equation}
with Hamiltonian
\begin{equation}\label{basic_generator}
H=-\frac{1}{2}\Delta+V(x)+W(\epsi x).
\end{equation}
The electron moves in $\mathbb{R}^d$ and the solution to \eqref{basic_dynamics} defines the 
unitary time evolution $U^\epsi(t)\psi(x)=\mathrm{e}^{-itH}\psi(x)=\psi(x,t)$ in 
$L^2(\mathbb{R}^d)$. We have chosen units such that $\hbar=1$ and the mass of the particle 
$m=1$. $V(x)$ is a periodic potential with average lattice spacing $a$. The precise conditions 
on $V$ will be spelled out in the following section, where we also describe the direct fiber 
integral decomposition for periodic Schr\"{o}dinger operators. The lattice spacing $a$ 
defines the microscopic spatial scale. $W(\epsi x)$ is an external electrostatic 
potential with dimensionless scale parameter $\epsi$, $\epsi\ll 1$,
which means that $W$ is slowly varying on the scale of the lattice. For real metals the condition
of slow variation is satisfied even for the strongest external electrostatic fields
available, cf.\ \cite{ashcroft-mermin}, Chapter 13.

The external forces due to $W$ are of order $\epsi$ and therefore have to act over a 
time of order $\epsi^{-1}$ to produce finite changes, which defines the macroscopic
time scale. We will mostly work in the microscopic coordinates $(x,t)$ of 
(\ref{basic_dynamics}). For sake of comparison we note that the macroscopic space-time
scale $(x',t')$ is defined through $x=\epsi^{-1}x'$ and $t=\epsi^{-1}t'$.
With this scale change
Eqs.\ \eqref{basic_dynamics}, \eqref{basic_generator} read
\begin{equation} \label{scaled_dynamics}
\begin{array}{l} \displaystyle{
i\epsi\frac{\partial}{\partial t^\prime}\psi=
H\psi\,,}\\ \\ \displaystyle{
H=\left(-\epsi^2\frac{1}{2}\Delta^\prime
+V(x^\prime/\epsi)+W(x^\prime)\right)}
\end{array}
\end{equation}
with initial conditions 
$\psi^\epsi(x^\prime)=\epsi^{-d/2}\psi(x^\prime/\epsi)$. 
If $V=0$, Eq.\ \eqref{scaled_dynamics} is the usual semiclassical limit with 
$\epsi$ set equal to $\hbar$. Thus our problem is to understand how an additional 
periodic, but rapidly oscillating potential modifies the standard picture.

The two scale problem \eqref{basic_dynamics}, \eqref{basic_generator} 
can be attacked along several routes. 
A first choice would be time dependent WKB 
\cite{buslaev,buslaev-grigis,gerard-martinez-sjostrand,guillot-ralston-trubowitz}. 
In the limit $\epsi\to 0$, for each energy band separately, one obtains a 
Hamilton-Jacobi equation for the phase and a transport equation for the amplitude 
of the wave function $\psi(x,t)$. As a main draw-back of this method, generically, 
the solution to the Hamilton-Jacobi equation develops singularities after some finite 
macroscopic time. If $V=0$, it is well understood how to go beyond such caustics by 
introducing new coordinates on the Lagrangian manifold. For \eqref{basic_dynamics}, 
\eqref{basic_generator} a corresponding program has not yet been attempted. The 
results \cite{buslaev,buslaev-grigis,gerard-martinez-sjostrand,guillot-ralston-trubowitz} 
are valid only over a finite macroscopic time span with a duration depending on the 
initial wave function.

Another variant is to establish the semiclassical limit through the convergence of Wigner 
functions. In our context one defines a band Wigner function 
$W^\epsi_n(r,k,t)$ depending on the band index $n$ and as a function of the position 
and quasimomentum. One then wants to prove that in the limit $\epsi\to 0$ 
$W^\epsi_n(t)$ converges to $\overline{W}_n(t)$, which is the initial band Wigner 
function $\overline{W}_n(0)$ evolved according to the semiclassical flow 
\eqref{semiclassical_dynamics}. Such a result is established in 
\cite{gerard-markowich-mauser-poupaud,markowich-mauser-poupaud} for the case of zero 
external potential, the general case being left open as a challenging problem.

A third approach to the semiclassical limit for $V=0$ is the strong convergence of 
Heisenberg operators \cite{asch-knauf,avron-seiler-yaffe,nenciu,spohn}. 
We briefly recall its main features. We define, as unbounded operators on $L^2(\mathbb{R}^d)$,
\begin{eqnarray*}
&x(t):=\mathrm{e}^{itH}x\mathrm{e}^{-itH},\\ \\
&p(t):=\mathrm{e}^{itH}p\mathrm{e}^{-itH},\quad p=-i\nabla_x,
\end{eqnarray*}
where $H$ is the Hamiltonian in \eqref{basic_generator} with $V=0$. The goal is to establish 
the strong limit of 
$$x^\epsi(t)\psi=\epsi x(\epsi^{-1}t)\psi,\quad 
p^\epsi(t)\psi=p(\epsi^{-1}t)\psi$$ 
as $\epsi\to 0$ with $\psi$ in a suitable domain. In the trivial case of free 
motion, $W=0$, this amounts to the strong convergence of 
$x^\epsi(t)\psi=(\epsi x+pt)\psi$, $p^\epsi(t)\psi=p\psi$, which 
yields $\lim_{\epsi\to 0}x^\epsi(t)=pt$, 
$\lim_{\epsi\to 0}p^\epsi(t)=p$. The general case requires more work 
\cite{robert}. One obtains the strong limits
\begin{equation} \label{semiclassical_limit}
\begin{array}{l} \displaystyle{
\lim_{\epsi\to 0}x^\epsi(t)=r(p,t)\,,}\\ \\ \displaystyle{
\lim_{\epsi\to 0}p^\epsi(t)=u(p,t)\,.}
\end{array}
\end{equation}
Here $r(p,t)$, $u(p,t)$ are solutions of
\begin{equation}\label{semiclassical_flow}
\dot{r}=u,\quad \dot{u}=-\nabla W(r)
\end{equation}
with initial conditions $r_0=0$, $u_0=p$. The initial condition $r_0=0$ reflects that 
$\vert\psi\vert^2$  looks like $\delta(r)$ on the macroscopic 
scale, provided that $\Vert\psi\Vert_2=1$. For general initial conditions, 
$r_0\ne 0$, we would have to shift the initial 
$\psi$ by $\epsi^{-1}r_0$.

The strong operator convergence may look slightly abstract, but all the desired physical 
information can be deduced. E.g., the initial $\psi$ defines the momentum distribution 
$\vert\widehat{\psi}(k)\vert^2$ independent of $\epsi$ and the $\delta(r)$ spatial 
distribution in the limit $\epsi\to 0$. Then, according to \eqref{semiclassical_limit}, 
for small $\epsi$ the position distribution at time $t$ is given by
\begin{eqnarray*}
&\int_{\mathbb{R}^d}f(x)\vert\psi^\epsi(x,t)\vert^2\,dx=(\psi,f(x^\epsi(t))\psi)\\
&\simeq(\psi,f(r(p,t))\psi)=\int\vert\widehat{\psi}(k)\vert^2f(r(k,t))\,dk,
\end{eqnarray*}
which means that the phase space distribution $\delta(r)\vert\widehat{\psi}(k)\vert^2\,dr\,dk$ 
is transported according to the semiclassical flow \eqref{semiclassical_flow}. The spatial 
marginal of this distribution at time $t$ is the desired approximation to the true
position distribution  
$\vert\psi^\epsi(x,t)\vert^2$. $\vert\psi^\epsi(x,t)\vert^2$ may oscillate 
rapidly on small scales and some averaging, as embodied by the test function $f$, is needed.

In this paper we investigate the semiclassical limit \eqref{basic_dynamics}, 
\eqref{basic_generator} through the strong convergence of the position operator 
$x^\epsi(t)$. 
 We will show that, in the limit $\epsi\to 0$, 
$x^\epsi(t)$ is diagonal with respect to the band index and in each band the 
structure is analogous to \eqref{semiclassical_limit} with $p$ replaced by the 
quasimomentum $k$ and \eqref{semiclassical_flow} replaced by \eqref{semiclassical_dynamics}.
More generally we will consider the semiclassical limit of the Weyl quantized
operators $a^W(\epsi x,p)$, whose classical symbol is periodic in $p$.

To give a short outline: In the following section we collect some properties of periodic 
Schr\"{o}dinger operators. In Section \ref{sect_main_results} we state our main results, 
which are proved in Sections \ref{effective}, 
\ref{sect_asymptotic_position}, and \ref{PsiDO}, respectively. 
In Section \ref{explain_results} we discuss some implications for the position and quasimomentum
distributions, and, more generally, for the band Wigner functions.
The difficulties arising from band crossings are explained in Section \ref{outlook}.

\section{Periodic Schr\"odinger operators}\label{sec_zwo}

For the periodic potential $V$ we will need only some rather minimal assumptions, which we state as
\begin{Con}[$\mathrm{C_{per}}$]
Let $\Gamma\simeq\mathbb{Z}^d$ be the lattice generated by the basis 
$\{\gamma_1,\ldots,\gamma_d\}$, $\gamma_i\in\mathbb{R}^d$. Then $V(x+\gamma)=V(x)$ for 
all $x\in\mathbb{R}^d$, $\gamma\in\Gamma$. 
Furthermore, we assume $V$ to be infinitesimally operator bounded with respect to $H_0$.
The last condition is satisfied, e.g., if
$V\in L^p(M)$, where $M$ is the fundamental domain of\, $\Gamma$, and $p=2$ 
for $d\leq 3$ and $p>d/2$ for $d> 3$, respectively.
\end{Con}
\noindent ($\mathrm{C_{per}}$) will be assumed throughout.

We recall the {\em Bloch-Floquet} theory for the spectral representation of 
\begin{equation}\label{periodic_generator}
H_\mathrm{per}=\frac{1}{2}p^2+V(x)\,.
\end{equation}
The reciprocal lattice $\Gamma^*$ is defined as the lattice generated by the dual basis 
$\{\gamma_1^*,\ldots,\gamma_d^*\}$ determined by $\gamma_i\cdot\gamma_j^*=2\pi\delta_{ij}$, 
$i,j=1,\ldots,d$. The fundamental domain of $\Gamma$ is denoted by $M$, the one of $\Gamma^*$ 
by $M^*$. $M^*$ is usually referred to as {\em first Brillouin zone}. 
If we identify opposite edges of $M$, resp.\ $M^*$, then it becomes a flat 
$d$-torus denoted by $\mathbb{T}=\mathbb{R}^d/\Gamma$, resp.\ 
$\mathbb{T}^*=\mathbb{R}^d/\Gamma^*$.

Let us introduce the Bloch-Floquet transformation, which should be viewed as a 
discrete Fourier transform, through 
\[
(\mathcal{U}\psi)(k,x):=\sum_{\gamma\in\Gamma}
\mathrm{e}^{-i(x+\gamma)\cdot k}\psi(x+\gamma),\quad (k,x)\in\mathbb{R}^{2d}\,,
\] 
for $\psi\in\mathcal{S}(\mathbb{R}^d)$.
Clearly,
\begin{equation}\label{BC}
\begin{array}{l} \displaystyle{
(\mathcal{U}\psi)(k,x^\prime+\gamma)=(\mathcal{U}\psi)(k,x^\prime)\,,} \\ \\ 
\displaystyle{ 
(\mathcal{U}\psi)(k^\prime+\gamma^*,x)
=\mathrm{e}^{-ix\cdot \gamma^*}(\mathcal{U}\psi)(k^\prime,x)\,.}
\end{array}
\end{equation}
Therefore it suffices to specify $\mathcal{U}\psi$ on the set $M^*\times M$ and,
if needed, extend it to all of $\mathbb{R}^{2d}$ by (\ref{BC}). 
The linear map 
$\mathcal{U}:L^2(\mathbb{R}^d)\supset\mathcal{S}(\mathbb{R}^d)
\to\mathcal{H}:=\int^\oplus_{M^*}L^2(M)\,dk,$ with $dk$ the normalized Lebesgue measure on 
$M^*$, has norm one and can thus be extended to all of $L^2(\mathbb{R}^d)$ by continuity. 
$\mathcal{U}$ is surjective as can be seen from the inverse mapping 
$$(\mathcal{U}^{-1}\phi)(x):=\int_{M^*}\mathrm{e}^{ix\cdot k}\phi(k,x)\,dk,$$ 
which has norm one. Thus $\mathcal{U}:L^2(\mathbb{R}^d)\to\mathcal{H}$ is unitary. 

To transform $H_\mathrm{per}$ under $\mathcal{U}$, we first note that 
$\tilde p = \mathcal{U}p\mathcal{U}^{-1} = D_x + k$,
with $D_x = -i\nabla_x$. Therefore
\[
\tilde{H}_{\mathrm{per}}:=\mathcal{U}H_{\mathrm{per}}\mathcal{U}^{-1}=
\int_{M^*}^\oplus H_{\mathrm{per}}(k)\,dk\,,
\] 
and
\[
H_\mathrm{per}(k)=\frac{1}{2}(D_x+k)^2+V(x),\quad k\in\mathbb{R}^d\,.
\]
$H_\mathrm{per}(k)$ acts on $L^2(M)$ with $k$-independent domain $D:=H^2(\mathbb{T})$. 
 $\psi\in D$ 
is periodic in $x$. $H_\mathrm{per}(k)$ is a semi-bounded self-adjoint operator, since by 
condition ($\mathrm{C_{per}}$) $V$ is infinitesimally operator bounded with respect to 
$-\Delta$ \cite{cycon-froese-kirsch-simon}. 
In particular, $H_{\mathrm{per}}(k)$ is an entire 
analytic family of type (B) in the sense of Kato for $k\in\mathbb{C}^d$. Since the resolvent 
of $H_0(k)=\frac{1}{2}(D_x+k)^2$ is compact, the resolvent 
$R_\lambda(H_{\mathrm{per}}(k)):=(H_{\mathrm{per}}(k)-\lambda)^{-1}$, 
$\lambda\ne\sigma(H_{\mathrm{per}}(k))$, is also compact, and $H_{\mathrm{per}}(k)$ has a 
complete set of (normalized) eigenfunctions
$\varphi_n(k)\in H^2(\mathbb{T})$, $n\in\mathbb{N}$, called  {\em Bloch functions}.  
The corresponding eigenvalues $E_n(k)$, 
$n\in\mathbb{N}$, accumulate at infinity and we enumerate them according to their magnitude 
and multiplicity, $E_1(k)\leq E_2(k)\leq\ldots$\ . $E_n(k)$ is called the $n$-th 
{\em band function}. We note that 
$H_\mathrm{per}(k)=\mathrm{e}^{-ix\cdot\gamma^*}
H_\mathrm{per}(k+\gamma^*)\mathrm{e}^{ix\cdot\gamma^*}$. Therefore $E_n(k)$ is periodic with 
respect to $\Gamma^*$. If $E_{n-1}(k)<E_n(k)<E_{n+1}(k)$ for all $k\in M^*$ (in particular 
$E_n(k)$ is nondegenerate), then the $n$-th 
band is {\em isolated}. In this case $E_n$ and the corresponding projection operator are real 
analytic functions as a consequence of analytic perturbation theory \cite{kato}.
We denote by $\mathcal{I}\subset\mathbb{N}$ the set of indices of isolated bands.

It will be convenient to have also a notation for the spectral subspaces. Let 
$P_n(k):L^2(M)\to L^2(M)$ denote the orthogonal projection onto the $n$-th eigenspace of 
$H_\mathrm{per}(k)$. Similarly, we set $Q_n(k)=\boldsymbol{1} -P_n(k)$. Their direct fiber 
integral is denoted by 
\[
\tilde{P}_n=\int^\oplus_{M^*}P_n(k)\,dk\,.
\] 
$\tilde{P}_n$ projects 
onto the $n$-th band subspace in $\mathcal{H}$ and $P_n=\mathcal{U}^{-1}\tilde{P}_n\mathcal{U}$
 projects onto the $n$-th band subspace in $L^2(\mathbb{R}^d)$. We have
\begin{eqnarray}\label{coefficients}
(\tilde{P}_n\psi)(k,\cdot)&=&P_n(k)\psi(k,\cdot)=
(\varphi_n(k),\psi(k))_{L^2(M)}\varphi_n(k,\cdot) \nonumber\\ 
&=&\psi_n(k)\varphi_n(k,\cdot)\,.
\end{eqnarray}
The coefficient functions $\psi_n\in L^2(M^*)$ and are called the {\em Bloch coefficients} in 
the $n$-th band subspace. For the index set $\mathcal{I}\subset\mathbb{N}$ of isolated bands 
we set $P_\mathcal{I}=\sum_{n\in\mathcal{I}}P_n$. 

\begin{Rem}
To have a concise notation, we will use a tilde for operators acting on $\mathcal{H}$. Thus 
if $A$ is an operator on $L^2(\mathbb{R}^d)$, then $\tilde{A}=\mathcal{U}A\mathcal{U}^{-1}$. 
If $A$ has a direct fiber decomposition, then $\tilde{A}=\int_{M^*}^\oplus A(k)\,dk$ with 
$A(k)$ acting on the fiber $L^2(M)$ of $\mathcal{H}$.
\end{Rem}

\section{Main results}\label{sect_main_results}

For the potentials we assume $(\mathrm{C_{per}})$ for $V$ and in addition
\begin{Con}[$\mathrm{C_{ex}}$] The external potential $W\in\mathcal{S}(\mathbb{R}^d)$.
\end{Con}

To state the semiclassical limit, we first have to explain the classical dynamics which 
will serve as a comparison. For each $n\in\mathcal{I}$ the classical phase space is 
$\mathbb{R}^d\times\mathbb{T}^*$, where $\mathbb{T}^*=\mathbb{R}^d/\Gamma^*$. As $n$-th band 
Hamiltonian we have 
\[
h_n(r,k)=E_n(k)+W(r),\quad (r,k)\in\mathbb{R}^d\times\mathbb{T}^*\,,
\]
and the classical dynamics in the $n$-th band is governed by
\begin{equation}\label{dynamical_system}
\dot{r}_n=\nabla_kE_n(k_n),\quad \dot{k}_n=-\nabla_rW(r_n).
\end{equation}
Since we want to prove the strong convergence of the position operator, as in the case 
$V\equiv 0$, we have to lift \eqref{dynamical_system} to operators on 
$\mathcal{H}$. For this purpose we solve \eqref{dynamical_system} with initial condition 
$r_n(0)=0$, $k_n(0)=k$. We denote the solution by $(r_n(t;k),k_n(t;k))$, regarded as functions 
of $k\in\mathbb{T}^*$. For $\psi\in \mathcal{H}$, we define 
\[
(R(t) \psi)(k,x) = \sum_{n\in\mathcal{I}}  r_n(t;k) \tilde{P}_n\psi(k,x)\,,
\]
and analogously, for later use,
\[
(K(t) \psi)(k,x) = \sum_{n\in\mathcal{I}}  k_n(t;k) \tilde{P}_n\psi(k,x)\,.
\] 
\begin{The}\label{MT3}
Let the conditions $(\mathrm{C_{per}})$, $(\mathrm{C_{ex}})$ be satisfied. Let 
\[
x^\epsi(t) = \epsi U^\epsi(-t/\epsi)\,x\,U^\epsi(t/\epsi)
\,.
\] 
Then for every $\psi\in \mbox{Ran} P_{\mathcal{I}}\cap D(\vert x\vert)\cap H^2$, 
with $H^2$ the second Sobolev
space,   
\[
\lim_{\epsi\to 0}x^\epsi(t)\psi=
\mathcal{U}^{-1}R(t)\mathcal{U}\psi
\]
strongly.
\end{The}
Theorem \ref{MT3} will be proved in several steps. First we show that
in the semiclassical limit transitions from and to isolated band subspaces are suppressed on
the level of the unitary groups. We define 
$H^n_\mathrm{diag}=P_nHP_n+Q_nHQ_n$ and 
$U^{\epsi,n}_\mathrm{diag}(t):=\exp(-itH^n_\mathrm{diag})$. In Section \ref{effective} 
we will prove
\begin{The}\label{MT1}
For any $n\in\mathcal{I}$ we have 
$$\lim_{\epsi\to 0}\left(U^\epsi(t/\epsi)-U^{\epsi,n}_\mathrm{diag}
(t/\epsi)\right)=0$$ 
in $B(H^1,L^2)$, where $H^1$ is the first Sobolev space.
\end{The}

The position operator is not diagonal with respect to the $n$-th band subspace and
we define its diagonal part  by 
$x^n_\mathrm{diag}=P_nxP_n+Q_n x Q_n$
with the time evolution 
\[
x^{\epsi,n}_\mathrm{diag}(t):=
\epsi U^{\epsi,n}_\mathrm{diag}(-t/\epsi)x^n_\mathrm{diag}
U^{\epsi,n}_\mathrm{diag}(t/\epsi)\,. 
\]
Our second step is to prove that the off-diagonal part of $x^\epsi(t)$
vanishes in the limit $\epsi\to 0$.
\begin{The}\label{MT2}
For $n \in \mathcal{I}$ 
\begin{equation} \label{MT2_eq}
\lim_{\epsi\to 0}  \left(x^\epsi(t)-x^{\epsi,n}_\mathrm{diag}(t)
\right)=0
\end{equation}
in $B(H^2,L^2)$.
\end{The}
By construction we have $[x^{\epsi,n}_{\rm diag}(t), P_n]=0$ and it
suffices to study the dynamics in the $n$-th band subspace. This subspace is isomorphic
to $L^2(\mathbb{T}^*)$ and, up to errors of higher order,
$x^{\epsi,n}_{\rm diag}(t)$ can be replaced by $x^{\epsi,n}_{\rm sc}(t)$
whose time evolution is governed by a Hamiltonian of the form
\[
\widetilde H^{\epsi,n}_{\rm sc} = E_n(k) + W(-i\epsi\nabla_k)\,.
\]
At this stage we can apply the standard machinery of semiclassics, except that
formally the roles of position and momentum have been interchanged and the
new position space is the flat torus rather than $\mathbb{R}^d$.

So far we focused on the position operator, since the electronic density is the
most accessible quantity experimentally and it corresponds in essence to a 
suitable function of the position. On more general grounds one would like to characterize a wider
class of semiclassical observables.  
 One further obvious candidate is the momentum $p$.
In the Bloch-Floquet basis we have $\widetilde p= k+D_x$. $k$ is semiclassical, as
being canonically conjugate to $-i\nabla_k$:
\begin{The} \label{MT5}
 Let 
\[
k^\epsi(t) =  U^\epsi(-t/\epsi)\,\mathcal{U}^{-1}\,k\,\mathcal{U}\,U^\epsi(t/\epsi)
\,.
\] 
Then for every $\psi\in \mbox{Ran}P_\mathcal{I}$
\begin{equation} \label{MT5_eq}
\lim_{\epsi\to 0}k^\epsi(t)\psi=
\mathcal{U}^{-1}K(t)\mathcal{U}\psi
\end{equation}
strongly.
\end{The}
\noindent On the other hand, $D_x$ is unbalanced because there is no
extra factor of $\epsi$. Thus $p(t/\epsi)$ has a limit only when averaged over
time (compare with Section \ref{sect_asymptotic_position}). 

It is relatively easy to see (cf.\ Section \ref{MT4S}) that Theorems \ref{MT3} and \ref{MT5}
imply the semiclassical limit also for bounded functions of $x^\epsi(t)$ resp.\ of $k^\epsi(t)$
(cf.\ Lemma \ref{fl}).
Next note that for
$\Gamma^*$-periodic functions $g$,
$g(\cdot+\gamma^*)= g(\cdot)$ for all $\gamma^*\in\Gamma^*$, we have $\mathcal{U}g(p)
\mathcal{U}^{-1}=g(k)$ and hence, by the functional calculus for self-adjoint operators, 
$g(p^\epsi(t))=g(k^\epsi(t))$. 
Therefore we introduce the set 
$\mathcal{O}(0)\subset  C(\mathbb{R}^d\times
\mathbb{R}^d, \mathbb{R}) $ of bounded and continuous semiclassical symbols 
which vanish if the first argument approaches infinity and
are $\Gamma^*$-periodic in their second
argument. For $a\in \mathcal{O}(0)$ we introduce its Weyl quantization
\begin{equation} \label{weyl}
(a^{\rm W}\psi)(x)=\frac{1}{(2\pi)^d}\int a\left( \frac{x+y}{2},\xi\right)
e^{i(x-y)\cdot \xi}\psi(y)\,d\xi dy
\end{equation}
as a bounded operator on $L^2(\mathbb{R}^d)$. 
The operator corresponding to the symbol $a(\epsi x,\xi)$ will be denoted by
$a^{{\rm W},\epsi}$ and we set, as before,
\begin{equation}
a^{{\rm W},\epsi}(t)=U^{\epsi}(-t/\epsi) a^{{\rm W},\epsi}
U^{\epsi}(t/\epsi)\,.
\end{equation}
\begin{The} \label{MT4}
Let the conditions $(\mathrm{C_{per}})$, $(\mathrm{C_{ex}})$ be satisfied
and $a\in \mathcal{O}(0)$. Then for every $\psi \in P_\mathcal{I}L^2$ we have
\[
\lim_{\epsi\to 0} a^{{\rm W},\epsi}(t)\psi =
\mathcal{U}^{-1}a(R(t),K(t))\mathcal{U}\psi\,.
\]
\end{The}

\section{Semiclassical distributions} \label{explain_results}

Theorems \ref{MT3} and \ref{MT4} tell us how the quantum distributions
behave in the semiclassical limit. Let us first consider
the initial $\psi\in P_{\mathcal{I}}\mathcal{H}$. Its scaled position distribution
is $\epsi^{-d}|\psi(x/\epsi)|^2$ which converges to
$\delta(x)$ as a measure. The quasimomentum distribution
$\sum_{n\in\mathcal{I}} |\psi_n(k)|^2$ is independent of $\epsi$. Thus it is natural to
chose
\begin{equation}
\rho (dr\,dk) = \sum_{n\in\mathcal{I}} \delta(r)|\psi_n(k)|^2 \,dr\,dk\, =
\sum_{n\in\mathcal{I}} \rho_n(dr\,dk)
\end{equation}
as initial distribution for the semiclassical flow (\ref{dynamical_system}).
We could consider more general initial measures at the expense of making
$\psi$ itself $\epsi$-dependent. 
For example the shifted initial measure $\sum_{n\in \mathcal{I}}\delta(r-r_0)|\psi_n(k)|\,dr\,dk$
is approximated by $\psi(x-\epsi^{-1}r_0)$.
Under (\ref{dynamical_system}) 
$\rho(dr\,dk)$ evolves to $\rho(dr\,dk,t)= \sum_{n\in\mathcal{I}} \rho_n(dr\,dk,t)$.
Each $\rho_n$ satisfies weakly the transport equation
\begin{equation}
\frac{\partial}{\partial t}\rho_n = -\nabla E_n(k)\cdot \nabla_r\rho_n + \nabla V(r)\cdot
\nabla_k\rho_n
\end{equation}
with initial condition $\rho_n(dr\,dk,0)=\rho_n(dr\,dk)$. We define the position and quasimomentum
marginals through
\begin{equation}
\rho(dr,t)=\int_{M^*} \rho(dr\,dk,t)\,,\qquad \rho(dk,t)=\int_{\mathbb{R}^d}\rho(dr\,dk,t)\,.
\end{equation}
To connect with the quantum evolution we consider the quantum mechanical position distribution
\begin{equation}
\rho^\epsi(dx,t) = \epsi^{-d}|\psi(x/\epsi, t/\epsi)|^2 \,dx
\end{equation}
as a probability measure on $\mathbb{R}^d$. From Theorem \ref{MT3} and Lemma \ref{fl} we conclude that
\begin{equation}
\lim_{\epsi\to 0} f(x^\epsi(t))\psi = \mathcal{U}^{-1}f(R(t))\mathcal{U} \psi
\end{equation}
for $f\in C_\infty (\mathbb{R}^d)$. In particular,
\begin{equation} \label{46}
\lim_{\epsi\to 0} \int \rho^\epsi(dx,t)f(x) =
\lim_{\epsi\to 0} (\psi, f(x^\epsi(t))\psi) = 
(\mathcal{U} \psi, f(R(t))\mathcal{U} \psi)
\end{equation}
and we only have to compute the expression on the right hand side. Using that
\[
(\mathcal{U}\psi)(x,k) =  \sum_{n\in\mathcal{I}} \psi_n(k)\rho_n(x,k)
\]
we have
\begin{equation}
(\mathcal{U} \psi, f(R(t))\mathcal{U} \psi) =  \sum_{n\in\mathcal{I}} 
\int_{M^*} |\psi_n(k)|^2 f(r_n(t;k))\,dk=
 \sum_{n\in\mathcal{I}} \int_M \rho_n(dr,t)f(r)\,.
\end{equation}
Thus the positional distribution $\rho^\epsi(dx,t)$ converges weakly as a measure
to the incoherent sum $ \sum_{n\in\mathcal{I}} \rho_n(dr,t)$.

By the same reasoning, if $g$ is a $\Gamma^*$-periodic function,
then by Theorem \ref{MT5} and Lemma \ref{fl} 
\begin{equation}
\lim_{\epsi\to 0} g(p(t/\epsi))\psi = \mathcal{U}^{-1} g(K(t))\mathcal{U}\psi\,.
\end{equation}
Therefore, if $\rho^\epsi(k,t)\,dk$ denotes the spectral measure
for the quasimomentum operator at time $t/\epsi$, we have 
\begin{equation} \label{49}
\lim_{\epsi\to 0} \rho^\epsi(k,t)\,dk = \sum_{n\in\mathcal{I}} \rho_n(dk,t)
\end{equation}
weakly as measures.

More generally for $\psi\in L^2$ we define the scaled Wigner function by
\begin{equation}
W^\epsi(x,k,t) = \sum_{\gamma\in\Gamma}\epsi^{-d}\psi(\epsi^{-1}x-
\frac{1}{2}\gamma,\epsi^{-1}t) \psi^*(\epsi^{-1}x+
\frac{1}{2}\gamma,\epsi^{-1}t) 
e^{ik\cdot\gamma}
\end{equation}
with $x\in\mathbb{R}^d$, $k\in M^*$. We think of $W^\epsi$ as a signed,
bounded measure over $\mathbb{R}^d\times M^*$.
The Wigner function yields expectations of Weyl quantized operators through
\begin{equation}
\left( \psi, e^{iH^\epsi t/\epsi}a^{\rm W,\epsi}  e^{-iH^\epsi t/\epsi}\psi\right)
=
\int_{\R^d\times M^*}W^\epsi(x,k,t)a(x,k)\,dx\,dk
\end{equation}
with $a$ $\Gamma^*$-periodic in its second argument. From Theorem \ref{MT4}
we therefore deduce that
\begin{equation}
\lim_{\epsi\to 0} W^\epsi(r,k,t)\,dr\,dk \, = \rho(dr\,dk,t)
\end{equation}
weakly as measures. The limits (\ref{46}) and (\ref{49}) are the particular
cases, where either $a(x,k) = f(x)$ or $a(x,k)= g(k)$.

\section{Convergence of the unitary groups}\label{effective}

By definition, the time evolution generated by $H_\mathrm{per}$ leaves invariant 
the band subspaces Ran($P_n$)  for all $n\in\mathbb N$. However,
 $W^\epsi(x)=W(\epsi x)$ does not respect the Bloch
decomposition and it will induce transitions between different bands. 
Since $W^\epsi$ is of slow variation, we expect such transitions to have a small 
amplitude as stated in Theorem \ref{MT1}. 

$W^\epsi$ transforms under $\mathcal{U}$ as
\begin{eqnarray}
        (\mathcal{U}W^\epsi\psi)(k,x)&=&\sum_{\gamma\in\Gamma}\mathrm{e}^{-i(x+\gamma)
\cdot k}W(\epsi(x+\gamma))\psi(x+\gamma)\nonumber\\
        &=&\sum_{\gamma\in\Gamma}\mathrm{e}^{-i(x+\gamma)\cdot k}(2\pi)^{-d/2}
\int_{\mathbb{R}^d}\widehat{W}(p)\mathrm{e}^{i\epsi(x+\gamma)\cdot p}\,dp\,\psi(x+\gamma)
\nonumber\\
        &=&(2\pi)^{-d/2}\int_{\mathbb{R}^d}\widehat{W}(p)(\mathcal{U}\psi)(k-\epsi p,x)\,dp
\nonumber\\
        &=& (2\pi)^{-d/2}\int_{\mathbb{R}^d}\widehat{W}^\epsi(p)(\mathcal{U}\psi)(k-p,x)\,dp
=:(\tilde{W}^\epsi\mathcal{U}\psi)(k,x)\, ,\label{convolution}
\end{eqnarray}
where $\widehat{W}^\epsi(p)=\epsi^{-d}\widehat{W}(p/\epsi)$ and we 
adopt the 
quasiperiodic extension \eqref{BC}. Since $\widehat{W}\in\mathcal{S}(\mathbb{R}^d)$, 
the integral 
\eqref{convolution} is well-defined and $\tilde{W}^\epsi=
\mathcal{U}W^\epsi\mathcal{U}^{-1}$ acts  on $\mathcal{H}$ as convolution with 
$\widehat{W}^\epsi$ in the fiber parameter $k$. 
$\widehat{W}^\epsi$ approximates a Dirac delta in the limit $\epsi\to 0$ and the 
shift in \eqref{convolution} becomes the identity operator.

In the Bloch-Floquet representation the full Hamiltonian \eqref{basic_generator} becomes 
\[
(\tilde{H}\psi)(k,\cdot)=H_\mathrm{per}(k)\psi(k,\cdot)+
(\tilde{W}^\epsi\psi)(k,\cdot)\,.
\] 
We expect the diagonal part of 
$W^\epsi$ to be dominant with the off-diagonal piece as a small correction. For such a 
decomposition it turns out to be convenient to fix the index $n$ of an isolated band and to 
project along $P_n$ and its complement $Q_n=\boldsymbol{1}-P_n$.
For $n\in\mathcal{I}$ we define the diagonal part $H^n_\mathrm{diag}$ of $H$  
as 
\[
{H}^n_{\mathrm{diag}}={P}_n{H}{P}_n+
{Q}_n{H}{Q}_n\,,
\] 
and the off-diagonal part of the external potential as 
\[
{W}^{\epsi,n}_{\mathrm{od}}=
{Q}_n{W}^\epsi{P}_n+
{P}_n{W}^\epsi{Q}_n\,.
\] 
Then 
\[
{H}={H}^n_{\mathrm{diag}}+{W}^{\epsi,n}_\mathrm{od}=
({H}_\mathrm{per}+{W}^{\epsi,n}_\mathrm{diag})+
{W}^{\epsi,n}_\mathrm{od}\,.
\]
We note that ${W}^{\epsi,n}_\mathrm{diag}$ and 
${W}^{\epsi,n}_\mathrm{od}$ are bounded operators and set 
\[
{U}^\epsi(t)=\mathrm{e}^{-it{H}},\quad
{U}^{\epsi,n}_{\mathrm{diag}}(t)=\mathrm{e}^{-it{H}^n_{\mathrm{diag}}}\,.
\]

To prove Theorem \ref{MT1} 
 we start by writing the difference of the two unitary groups 
in the Bloch representation as
\begin{equation}\label{difference_1}
\tilde{U}^\epsi(t/\epsi)-
\tilde{U}^{\epsi,n}_\mathrm{diag}(t/\epsi) = -\,i\epsi\int_0^{t/\epsi}\tilde{U}^\epsi(\epsi^{-1}t-s)
\left(\epsi^{-1}\tilde{W}^{\epsi,n}_{\rm od}\right)
\tilde{U}^{\epsi,n}_\mathrm{diag}(s)\,ds
\end{equation}
and we have to investigate the operator $\tilde{W}^{\epsi,n}_{\rm od}$.
 By definition, for 
$\psi\in\mathcal{H}$, we have 
\[
(\tilde{Q}_n\tilde{W}^\epsi\tilde{P}_n\psi)(k)=(2\pi)^{-d/2}\int_{\mathbb{R}^d}
\widehat{W}^\epsi(p)Q_n(k) P_n(k-p)\psi(k-p)\,dp\,,
\]
which vanishes strongly in the limit $\epsi\to 0$, since $\widehat{W}^\epsi$ 
localizes around $p=0$. 
To control the long times in \eqref{difference_1} we need uniform convergence of order $o(\epsi)$,
however. To have a more detailed information on $W^{\epsi, n}_{\rm od}$ we Taylor
expand of $P_n(k-p)$ around $P_n(k)$, leading, as we will show, to
\begin{equation} \label{expansion}
(\tilde{Q}_n\tilde{W}^\epsi\tilde{P}_n\psi)(k) = -\epsi(2\pi)^{-d/2}\int_{\mathbb{R}^d}
\widehat F^\epsi(p) Q_n(k)\nabla_k P_n(k) \psi(k-p) \,dp\,+o(\epsi)\,.
\end{equation}
Here $\widehat F^\epsi(p) := \widehat W^\epsi(p)\frac{p}{\epsi}$ is the Fourier
transform of $F^\epsi(x) = (D_xW)(\epsi x)$ and we will associate to $\widehat F^\epsi$ the operator 
$\tilde F^\epsi$ as in the case of $\widehat W^\epsi$,
\[
(\tilde{F}^\epsi\psi)(k)=(2\pi)^{-d/2}
\int_{\mathbb{R}^d}\widehat{F}^\epsi(p)\psi(k-p)\,dp\,.
\]
To justify \eqref{expansion} we first need to calculate $\nabla_k P_n(k)$. 
\begin{Lem}\label{smooth_projection}
Let $n\in\mathcal{I}$. Then 
\begin{eqnarray}\label{gradient_projection}
 \nabla_k P_n(k)&=&-\,Q_n(k)R_{E_n(k)}(H_\mathrm{per}(k))(D_x+k)P_n(k)\, \nonumber\\
&& -\, P_n(k)(D_x+k)R_{E_n(k)}(H_\mathrm{per}(k))Q_n(k)\,,
\end{eqnarray}
where $R_\lambda(H)= (H-\lambda)^{-1}$ is the resolvent of $H$. Thus
$P_n(\cdot)\in C^\infty(M^*;B(L^2(M)))$.
\end{Lem}
\begin{proof}
Using contour integrals we write 
\[
\nabla_k P_n(k)=-\, \frac{1}{2\pi i}\oint_{c_n(k)}\nabla_k 
R_\lambda(H_\mathrm{per}(k))\,d\lambda\,,
\]
where $c_n(k)$ is a closed rectifiable curve in the complex spectral plane which encircles 
$E_n(k)$ only. From
\begin{eqnarray*}
     0&=&\nabla_k\boldsymbol{1}=\nabla_k(H_\mathrm{per}(k)-\lambda)
R_\lambda(H_\mathrm{per}(k))\\
     &=&(D_x+k)R_\lambda(H_\mathrm{per}(k))+(H_\mathrm{per}(k)-\lambda)\nabla_k 
R_\lambda(H_\mathrm{per}(k))\,,
\end{eqnarray*}
we infer
\[
\nabla_k R_\lambda(H_\mathrm{per}(k))=-R_\lambda(H_\mathrm{per}(k))(D_x+k)
R_\lambda(H_\mathrm{per}(k))\,.
\]
Hence we get
\begin{eqnarray} \lefteqn{
     Q_n(k)\nabla_k P_n(k)\, =\, Q_n(k)\nabla_k P_n(k) (P_n(k) + Q_n(k))} \nonumber\\
     & = & \frac{1}{2\pi i}\oint_{c_n(k)}Q_n(k)R_\lambda(H_\mathrm{per}(k))(D_x+k)
R_\lambda(H_\mathrm{per}(k))P_n(k)\,d\lambda\nonumber\\
     & = & \frac{1}{2\pi i}\oint_{c_n(k)}R_\lambda(H_\mathrm{per}(k))Q_n(k)
\frac{1}{E_n(k)-\lambda}\,d\lambda\,(D_x+k)P_n(k)\nonumber\\
     & = & -R_{E_n(k)}(H_\mathrm{per}(k))Q_n(k)(D_x+k)P_n(k)\,,\label{offdiagonal_gradient}
\end{eqnarray}
where the term $Q_n(k)\nabla_k P_n(k) Q_n(k)$ vanishes, 
since in this case the integrand is an analytic function
on the whole interior of $c_n(k)$.
Note that $P_n(k)$ projects onto a subspace of finite energy, on which $D_x+k$ is bounded. 
The statement about continuity for this term  then follows 
from the continuity of $P_n(k)$, 
$E_n(k)$ and the assumption that $E_n(k)$ is isolated from the remainder of the spectrum.
An analogous computation for $P_n(k)\nabla_k P_n(k)$ leads to the second term 
in  (\ref{gradient_projection}).

Finally, $P_n(\cdot)\in C^\infty(M^*;B(L^2(M)))$ follows by induction. 
\end{proof}
From $Q_n(k) + P_n(k) = \boldsymbol{1}$ we conclude that $Q_n(k)$ is differentiable as well and
that $\nabla_k Q_n(k) = - \nabla_k P_n(k)$.

\begin{Lem}\label{first_order}
Let $n\in\mathcal{I}$. Then 
\[
\tilde{W}^{\epsi,n}_{\rm od} = - \epsi\left(\tilde Q_n\nabla_k\tilde P_n+
\tilde P_n \nabla_k\tilde Q_n\right)\cdot
\tilde F^\epsi + o(\epsi)
\] 
in $B(\mathcal{H},\mathcal{H})$, where $\nabla_k\tilde P_n := 
\int_{M^*}^\oplus\nabla_k P_n(k)\,dk$.
\end{Lem}
\begin{proof}
We will treat only the $\tilde Q_n\tilde W^\epsi \tilde P_n$
part of $\tilde W^{\epsi,n}_{\rm od}$ explicitly, since the argument for the second part 
is analogous.

Let $\psi\in\mathcal{H}$.
By Lemma \ref{smooth_projection} we are allowed to write the 
following well-defined identity, setting $e_p=p/\vert p\vert$,
\begin{eqnarray}
 \lefteqn{\int\widehat{W}^\epsi(p)P_n(k-p)\psi(k-p)\,dp} \nonumber\\
  &=&\int\widehat{W}^\epsi(p)\vert p\vert\left(\vert 
          p\vert^{-1}(P_n(k-p)-P_n(k))+e_p
         \cdot\nabla_k P_n(k)\right)\psi(k-p)\,dp\nonumber\\
  &&+\,P_n(k)\int\widehat{W}^\epsi(p)\psi(k-p)\,dp-\int\widehat{W}^\epsi(p)p
\cdot\nabla_k P_n(k)\psi(k-p)\,dp\nonumber\\ 
     &=&\epsi\int\widehat{W}^\epsi(p)\frac{\vert p\vert}{\epsi}
\left(\frac{P_n(k-p)-P_n(k)}{\vert p\vert}+e_p\cdot\nabla_k P_n(k)\right)\psi(k-p)\,dp
\label{oans}\\
     &&+\,(2\pi)^{d/2}P_n(k)(\tilde{W}^\epsi\psi)(k)-\epsi(2\pi)^{d/2}
\nabla_k P_n(k)\cdot(\tilde{F}^\epsi\psi)(k)\,.\label{zwoa}
\end{eqnarray}
If in \eqref{oans}, 
\eqref{zwoa} we apply $Q_n(k)$ from the left, the first term of \eqref{zwoa} vanishes and it 
remains to show that \eqref{oans}, divided by $\epsi$, tends to zero uniformly for all 
$\psi\in\mathcal{H}$. 

We split the integral into two parts. Let $R>0$ be arbitrary, 
$B_R=\{p\mid\vert p\vert\leq R\}$. We start with
\begin{eqnarray*}
\lefteqn{\int\limits_{M^*} \left\| \,\int\limits_{B_R}\widehat{W}^\epsi(p)
\frac{\vert p\vert}{\epsi}\Bigg(\frac{P_n(k-p)-P_n(k)}{\vert p\vert}\right.} \\ 
& & \left. \hspace{4cm}+\,
e_p\cdot\nabla_k P_n(k)\Bigg)\psi(k-p)\,dp \right\|_{L^2(M)}\,dk \\
 &\leq& \sup_{k\in M^*}\sup_{p\in B_R}\left\|| p|^{-1}\left(P_n(k-p)-P_n(k)\right)
+\,\,e_p\cdot\nabla_k P_n(k)\right\|\\
& &\hspace{2cm}\times\,\int_{B_R}\left|\widehat{W}^\epsi(p)\right|\frac{| p|}{\epsi}
\int_{M^*}\left\|\psi(k-p)\right\|_{L^2(M)}\,dk\,dp\\
     &\leq & \left\| \psi\right\|_\mathcal{H} \big\|\widehat{F}^\epsi\big\|_{L^1} \\
&&\hspace{1cm}\times \,
\sup_{k\in M^*,\,p\in B_R}\left\|| p|^{-1}(P_n(k-p)-P_n(k))+e_p\cdot\nabla_k P_n(k)\right\|\,.
\end{eqnarray*}
Since $\Vert\widehat{F}^\epsi\Vert_{L^1}$ does not depend on $\epsi$ and
since the difference quotient approaches the derivative uniformly on the compact domain
$M^*$, the $\mathcal{H}$-norm of the first part tends to zero uniformly.
For the remaining part we have
\begin{eqnarray*} \lefteqn{
\int\limits_{M^*}  \left\|\,\int\limits_{|p|>R}\widehat{W}^\epsi(p)\frac{| p|}{\epsi}
\Bigg(\frac{P_n(k-p)-P_n(k)}{| p|} \right.}\\ && \left. \hspace{4cm}+ \,\,
e_p\cdot\nabla_k P_n(k)\Bigg)\psi(k-p)\,dp\right\|_{L^2(M)} \, dk \\
     &\leq &  \left\|\psi\right\|_{\mathcal{H}} \big\| \widehat{F}^\epsi\big\|_{L^1(B_R^c)} 
\\
  &   &\hspace{1cm} \times\,  
\sup_{k\in M^*,\,p\in\mathbb{R}^d}\left\| | p|^{-1}(P_n(k-p)-P_n(k))+e_p\cdot\nabla_k P_n(k)
\right\| \,,
\end{eqnarray*}
which tends to zero uniformly as $\epsi\to 0$, since 
$\Vert\widehat{F}^\epsi\Vert_{L^1(B_R^c)}\to 0$ for any fixed $R>0$.
\end{proof}

As a consequence of Lemma \ref{first_order} the difference of the two unitary groups in 
Eq.\ \eqref{difference_1} can be written as
\begin{eqnarray} \label{difference_2}
   \lefteqn{ \tilde{U}^\epsi(t/\epsi)-\tilde{U}^{\epsi,n}_\mathrm{diag}
(t/\epsi)} \nonumber\\
   &  = \,i\epsi\int_0^{t/\epsi}\tilde{U}^\epsi(\epsi^{-1}t-s)
\left( \tilde{Q}_n\nabla_k\tilde{P}_n +  \tilde{P}_n\nabla_k\tilde{Q}_n \right)
          \cdot\tilde{F}^\epsi\,
\tilde{U}^{\epsi,n}_\mathrm{diag}(s)\,ds+o(1)\,.
\end{eqnarray}
We have to estimate the integral without losing one order of $\epsi$ from the 
integration over time.
As in the proof in \cite{avron-elgart} of the adiabatic theorem the idea is to rewrite the 
integrand as a time derivative, i.e.\ as a commutator of $\tilde{H}^n_\mathrm{diag}$ with 
an appropriately chosen operator $A$, at least up to an unavoidable error $o(1)$. 
 
Let us define for $n\in\mathcal{I}$ 
$$B_n(k)=R^2_{E_n(k)}(H_\mathrm{per}(k))Q_n(k)(D_x+k)P_n(k)\,.$$ 
\begin{Lem}
For $n\in\mathcal{I}$ we have 
\[
\tilde{Q}_n\nabla_k\tilde{P}_n + \tilde{P}_n\nabla_k\tilde{Q}_n
=[\tilde{B}_n+\tilde B^*_n,
\tilde{H}_\mathrm{per}]\,.
\]
\end{Lem}
\begin{proof}
Using the spectral decomposition and recalling
\[
Q_n(k)\nabla_k P_n(k)=-R_{E_n(k)}(H_\mathrm{per}(k))Q_n(k)(D_x+k)P_n(k)
\]
from  Lemma \ref{smooth_projection}, one directly computes
\begin{eqnarray*} \lefteqn{
B_n(k)H_\mathrm{per}(k)-H_\mathrm{per}(k)B_n(k)}\\
&=&-(H_\mathrm{per}(k)-E_n(k))R^2_{E_n(k)}(H_\mathrm{per}(k))Q_n(k)(D_x+k)P_n(k)\\
&=&-R_{E_n(k)}(H_\mathrm{per}(k))Q_n(k)(D_x+k)P_n(k)\\
&=&Q_n(k)\nabla_k P_n(k)\,.
\end{eqnarray*}
The lemma then follows from $\tilde{P}_n\nabla_k\tilde{Q}_n = 
- (\tilde{Q}_n\nabla_k\tilde{P}_n)^*$.
\end{proof}

\begin{Lem} \label{commu}
$\left[B_n + B_n^* 
,\tilde{W}^{\epsi,n}_\mathrm{diag}\right]\to 0$ in $B(\mathcal{H},\mathcal{H})$ 
as $\epsi$ tends to zero.
\end{Lem}
\begin{proof}
To have a concise notation in the following, expressions like 
$\tilde{W}^{\epsi,n}_\mathrm{diag}P_n(k)$ are understood in the sense that 
$\tilde{W}^{\epsi,n}_\mathrm{diag}$ acts on all $k$-depend\-ing objects on its right 
hand side. We recall that $\tilde{W}^{\epsi,n}_\mathrm{diag}=
\tilde{P}_n\tilde{W}^\epsi\tilde{P}_n+\tilde{Q}_n\tilde{W}^\epsi\tilde{Q}_n$. Hence
\begin{eqnarray*}
\lefteqn{\Big[ B_n(k), \tilde{W}^{\epsi,n}_\mathrm{diag}\Big] }\\
     &=&  Q_n(k)\Big[R^2_{E_n(k)}(H_\mathrm{per}(k))Q_n(k)(D_x+k)P_n(k),
\tilde{W}^{\epsi}\Big]P_n(k)\,.
\end{eqnarray*}
We now examine the commutators  $[P_n(k),\tilde{W}^\epsi]$,  
$[D_x+k,\tilde{W}^\epsi]$ and $[R^2_{E_n(k)}Q_n(k), \tilde{W}^\epsi]$ one by one.
It follows from the proof of Lemma \ref{first_order} that $[P_n(k),\tilde{W}^\epsi]$
vanishes as $\epsi\to 0$ and the analogous statement for 
$[R^2_{E_n(k)}Q_n(k), \tilde{W}^\epsi]$ can be shown to hold by a similar argument.
 Thus it remains to discuss the 
commutator $[D_x+k,\tilde{W}^\epsi]$. For $\psi\in H^1(\mathbb{R}^d)$ we compute 
\begin{eqnarray*} \lefteqn{
     (2\pi)^{d/2}([D_x+k,\tilde{W}^\epsi]\mathcal{U}\psi)(k)} \\
     &=&\int_{\mathbb{R}^d}\widehat{W}^\epsi(p)(((D_x+k)-(D_x+k-p))\mathcal{U}\psi)(k-p)\,dp\\
     &=&\epsi\int_{\mathbb{R}^d}\widehat{W}^\epsi(p)\epsi^{-1}p(\mathcal{U}\psi)
(k-p)\,dp\\
     &=&\epsi(\tilde{F}^\epsi\mathcal{U}\psi)(k)\,,
\end{eqnarray*}
which clearly vanishes uniformly for $\psi \in L^2$ as $\epsi\to 0$, 
since $F\in\mathcal{S}(\mathbb{R}^d,\mathbb{R}^d)$.
\end{proof}

In summary we have shown that 
\[
\left( \tilde{Q}_n\nabla_k\tilde{P}_n + \tilde{P}_n\nabla_k\tilde{Q}_n \right)
\cdot\tilde{F}^\epsi
=\left(\left[\tilde{B}_n + \tilde B_n^*,
\tilde{H}^n_\mathrm{diag}\right]+o(1)\right)\cdot \tilde{F}^\epsi\,,
\]
and it 
remains to check
\begin{Lem}
$\left[\tilde{H}^n_\mathrm{diag},\tilde{F}^\epsi\right]\to 0$ 
in $B(\mathcal{U}H^1, \mathcal{H})$ as 
$\epsi$ tends to zero.
\end{Lem}
\begin{proof}
The commutator 
\[
[H_\mathrm{per},F^\epsi] = -\frac{1}{2}\epsi^2 (\Delta F^\epsi)
-\frac{1}{2}\epsi (\nabla F^\epsi)\cdot\nabla-\frac{1}{2}\epsi
(\nabla\cdot F^\epsi)\nabla
\] 
vanishes  in $B(H^1,L^2)$  as $\epsi\to 0$. The commutator
$[\tilde{W}^{\epsi,n}_\mathrm{diag},\tilde{F}^\epsi]$
vanishes in $B(\mathcal{H},\mathcal{H})$, since the commutator of $\tilde{P}_n$ and 
$\tilde{Q}_n$ with  $\tilde{F}^\epsi$ are both of uniform 
order $o(1)$ (in $B(\mathcal{H},\mathcal{H})$) 
and $[\tilde{W}^\epsi,\tilde{F}^\epsi]$ vanishes identically.
\end{proof}

Defining 
\[
\tilde A_n = \left(\tilde B_n +\tilde B_n^*\right) \cdot \tilde F^\epsi\,,
\]
it follows that the integrand in (\ref{difference_2}) can be written as
\[
\left(\tilde{Q}_n\nabla_k\tilde{P}_n + \tilde{P}_n\nabla_k\tilde{Q}_n \right)
\cdot\tilde{F}^\epsi =\left[\tilde{A}_n,
\tilde{H}^n_\mathrm{diag}\right]+o(1)\,,
\]
where $o(1)$ is in the norm of $B(\mathcal{U}H^1,\mathcal{H})$.
(Note that for $A^\epsi\in B(L^2,L^2)$,
$\lim_{\epsi\to 0}$ $A^{\epsi}=0$ in $B(L^2,L^2)$ implies, in particular, that
also $\lim_{\epsi\to 0}A^{\epsi}=0$ in $B(H^1,L^2)$).
 
We are now ready for the
\begin{proof}[Proof of Theorem \ref{MT1}]
Since ${U}^{\epsi,n}_\mathrm{diag}(t):H^1\to H^1$ is bounded uniformly
in $t$ and $\epsi$ (cf.\ Section \ref{sect_asymptotic_position}), we obtain for
the difference \eqref{difference_2} of the unitary groups,
\begin{eqnarray} \label{last_difference} \displaystyle{
 \lefteqn{\Big(\tilde{U}^\epsi(t/\epsi)-\tilde{U}^{\epsi,n}_\mathrm{diag}
(t/\epsi)\Big)}}
 \nonumber\\
     & \displaystyle{
=\,-\,i\epsi\int_0^{t/\epsi}\tilde{U}^\epsi(\epsi^{-1}t-s)
\left[\tilde{A}_n,\tilde{H}^n_\mathrm{diag}\right]
\tilde{U}^{\epsi,n}_\mathrm{diag}(s) \,ds+o(1)\,.}
\end{eqnarray}
Abbreviating $X^n(s)=\tilde{U}^\epsi(-s)\tilde{U}^{\epsi,n}_\mathrm{diag}(s)$ and 
$\tilde{A}_n(s) = \tilde{U}^{\epsi,n}_\mathrm{diag}(-s)  \tilde{A}_n  
\tilde{U}^{\epsi,n}_\mathrm{diag}(s)$, we get,
using partial integration in \eqref{last_difference},
\begin{eqnarray*}\lefteqn{\hspace{-1cm}
     -i\epsi\int_0^{t/\epsi}\tilde{U}^\epsi(t/\epsi)\,X^n(s)\,
\tilde{U}^{\epsi,n}_\mathrm{diag}(-s)\left[\tilde{A}_n,\tilde{H}^n_\mathrm{diag}\right]
\tilde{U}^{\epsi,n}_\mathrm{diag}(s)\,ds}\\
     &=&\epsi\,\tilde{U}^\epsi(t/\epsi)\int_0^{t/\epsi}\,X^n(s)
\left(\frac{d}{ds}\tilde{A}_n(s)\right)\,ds\\
     &=&\epsi\left(\tilde A_n\,\tilde{U}^{\epsi,n}_\mathrm{diag}(t/\epsi)-
\tilde{U}^\epsi(t/\epsi)\,\tilde{A}_n\right)\\
     &&-\,\epsi\,\tilde{U}^\epsi(t/\epsi)\int_0^{t/\epsi}
\left(\frac{d}{ds}X^n(s)\right)\tilde{A}_n(s)\,ds\\
     &=&\epsi\left(\tilde{A}_n\,\tilde{U}^{\epsi,n}_\mathrm{diag}(t/\epsi)-
\tilde{U}^\epsi(t/\epsi)\,\tilde{A}_n\right)\\
     &&-\,i\epsi\,\tilde{U}^\epsi(t/\epsi)\int_0^{t/\epsi}\tilde{U}^\epsi(-s)\,
\tilde{W}^{\epsi,n}_{\rm od}\,\tilde{A}_n\,
\tilde{U}^{\epsi,n}_\mathrm{diag}(s)\,ds\,.
\end{eqnarray*}
For $\epsi\to 0$ the first term vanishes since $\tilde{A}_n$ is bounded and the second 
term vanishes, since $W^{\epsi,n}_{\rm od}$
tends to zero uniformly according to  Lemma \ref{first_order}.
\end{proof}

\section{Convergence of the position operator}\label{sect_asymptotic_position}

In this section we will study the asymptotics of the position operator $x^\epsi(t)$. 
As in the case of the unitaries we have to establish that the off-diagonal contributions to 
$x^\epsi(t)$ vanish in the limit $\epsi\to 0$.
\begin{proof}[Proof of Theorem \ref{MT2}]
Let $\psi\in  D(|x|)\cap H^2$ and $n\in \mathcal{I}$. Then
\begin{eqnarray}
\lefteqn{ \left\| \left(x^\epsi (t) - x^{\epsi,n}_{\rm diag} (t)\right) 
 \psi\right\|}\nonumber\\ \label{X1}
& \leq & 
\left\| \left(x^\epsi (t) - 
U_{\rm diag}^{\epsi,n} (-t/\epsi)\, 
x^\epsi\, U_{\rm diag}^{\epsi,n} (t/\epsi) 
\right) \psi\right\| \\ \label{X2}
&&+
\left\| \left( U_{\rm diag}^{\epsi,n} (-t/\epsi) 
\,x^\epsi\, U_{\rm diag}^{\epsi,n} (t/\epsi)
-  x^{\epsi,n}_{\rm diag} (t)\right) 
 \psi\right\|\,.
\end{eqnarray}
In order to estimate (\ref{X1}), note that we have 
\begin{equation}\label{mean_momentum}
x^\epsi(t)\psi=\epsi x\psi+\epsi\int_0^{t/\epsi}U^\epsi(-s)
D_xU^\epsi(s)\psi\,ds
\end{equation}
and
\begin{eqnarray*}\lefteqn{ 
U_{\rm diag}^{\epsi,n} (-t/\epsi)\, 
x^\epsi\, U_{\rm diag}^{\epsi,n} (t/\epsi) }\\
&=&
\epsi x \psi + \epsi \int_0^{t/\epsi}
U_{\rm diag}^{\epsi,n} (-s) \left( D_x + i \left[ W^{\epsi,n}_{\rm diag} , x \right]
\right)U_{\rm diag}^{\epsi,n} (s) \psi\,ds\\
&=&\epsi x \psi + \epsi \int_0^{t/\epsi}
U_{\rm diag}^{\epsi,n} (-s) D_x U_{\rm diag}^{\epsi,n} (s) \psi\,ds + o(1)\,.
\end{eqnarray*} 
The last equality holds,
since $[ W^{\epsi,n}_{\rm diag} , x] = o(1)$ in $B(L^2)$, as follows immediately from
the fact that $[ W^\epsi, P_n] = o(1)$ 
and $[ W^\epsi, Q_n] = o(1)$, cf.\ proof of Lemma \ref{first_order}.
Hence, using
(\ref{mean_momentum}), the remaining term from (\ref{X1}) is
\begin{eqnarray}\lefteqn{
\epsi\int_0^{t/\epsi}\left(U^\epsi(-s)D_xU^\epsi(s)-
U^{\epsi,n}_\mathrm{diag}(-s)D_xU^{\epsi,n}_\mathrm{diag}(s)\right)\psi\,
ds}\nonumber\\
&=&\int_0^{t}\left(U^\epsi(-s/\epsi)-
U^{\epsi,n}_\mathrm{diag}(-s/\epsi)
\right)D_x U^{\epsi,n}_\mathrm{diag}(s/\epsi)\psi\,ds
\label{zweiter_term}\\
&&+\,\int_0^{t}U^\epsi(-s/\epsi)D_x  
\left(U^\epsi(s/\epsi)-
U^{\epsi,n}_\mathrm{diag}(s/\epsi)\right) 
\psi\,ds\,.\label{dritter_term}
\end{eqnarray}
Using the fact that $V$ and $W$ are infinitesimally operator bounded
with respect to $-\frac{1}{2}\Delta$ and that $\psi\in H^2$, we get for 
$\psi(s):= U^{\epsi,n}_\mathrm{diag}(s/\epsi)\psi$
\begin{eqnarray*}
\left\|D_x^2  \psi(s) \right\| 
& \leq &
\left\|H^{\epsi,n}_{\rm diag} \psi(s)\right\|
+ \left\|(V+W^{\epsi,n}_{\rm diag})\psi(s)\right\| \\
& \leq &
\left\|H^{\epsi,n}_{\rm diag} \psi \right\|
+ c_1 \left\|D_x^2\psi(s)\right\| + c_2\left\|\psi\right\|\,,
\end{eqnarray*}
with $c_1<\frac{1}{2}$ and $c_2<\infty$.
Hence $\|D_x  U^{\epsi,n}_\mathrm{diag}(s/\epsi)\psi\|_{H^1}\leq c\|\psi\|_{H^2}$ with
$c$ independent of $s$ and $\epsi$ and we can apply
Theorem \ref{MT1} to conclude that the operator acting on $\psi$ in (\ref{zweiter_term}) 
vanishes in $B(H^2,L^2)$ as $\epsi\to 0$.

We come to (\ref{dritter_term}).
Let $\psi(s)=({U}^\epsi(s/\epsi)-
{U}^{\epsi,n}_\mathrm{diag}(s/\epsi))\psi$, then, by Cauchy-Schwarz,
\[
\left\| {D}_x\psi(s)\right\|^2 = \left(\psi(s),{D}^2_x\psi(s)\right)
\leq \left\|\psi(s)\right\| \, \left\|{D}^2_x
\psi(s)\right\|\,.
\]
The first factor tends to zero by Theorem \ref{MT1} whereas the second is uniformly bounded 
by the same argument as in the treatment of (\ref{zweiter_term}) a few lines above.

Next we rewrite (\ref{X2}) as
\[
\epsi U_{\rm diag}^{\epsi,n} (-t/\epsi) \,
x_{\rm od}^n\, U_{\rm diag}^{\epsi,n} (t/\epsi)
\]
with $x_{\rm od}^n := Q_n x P_n + P_n x Q_n$. This certainly vanishes 
as $\epsi\to 0$ if $x_{\rm od}^n$ can be shown to be a bounded operator. 
To see this, note that in Bloch representation $x$ acts as $i\nabla_k$.
Hence 
\[
(\mathcal{U} Q_n x P_n \psi)(k) = i Q_n(k) \nabla_k P_n(k)(\mathcal{U}\psi)(k) =
i Q_n(k) (\nabla_k P_n(k)) (\mathcal{U}\psi)(k)
\]
and thus $\|Q_n x P_n\| = \|\tilde Q_n \nabla_k\tilde P_n \|$.
Finally also $P_n x Q_n$ is bounded, since it is the adjoint of $Q_n x P_n$.
\end{proof}

\section{Semiclassical equations of motion for the position operator}\label{PsiDO}

As we have shown, on the macroscopic scale the position and quasimomentum operators 
commute with the projection on isolated bands. 
Thus it remains to investigate the semiclassical limit for each 
isolated band separately.
For this purpose we note that any $\psi \in \tilde P_n\mathcal{H}$ is of the
form $\psi_n(k)\varphi_n(x,k)$ with $\psi_n\in L^2(M^*)$. Since 
$\varphi_n$ already satisfies \eqref{BC}, we have to extend the
Bloch coefficients periodically. We determine now how
$H^{\epsi,n}_{\rm diag}$ acts on $L^2(M^*)$. We have $[H_{\rm per},\tilde P_n]=0$
and therefore $H_{\rm per}$ acts as multiplication by $E_n(k)$. For $W^{\epsi,n}_{\rm diag}$
we have  
\begin{eqnarray}\lefteqn{ \hspace{-.5cm}
\left( \tilde P_n \tilde W^\epsi \tilde P_n \mathcal{U} \psi \right) (k,x) }\nonumber \\
& = &(2\pi)^{-d/2}\int_{\mathbb{R}^d} 
\widehat W^\epsi (p) \big( \varphi_n(k),\varphi_n(k-p)\big)_{L^2(M)}
\psi_n(k-p) \,dp\,\varphi_n(k,x)\nonumber\\
& =: & (\tilde W^{\epsi,n}\psi_n)(k)\varphi_n(x,k)\,.
\end{eqnarray}
Thus $H^{\epsi,n}_{\rm diag}$ restricted to $\tilde P_n\mathcal{H}$ is unitarily equivalent to
$H^{\epsi,n} := E_n(k) + \tilde W^{\epsi,n}$.

To be able to use techniques from semiclassics we next approximate 
$\tilde W^{\epsi,n}$ by the operator $\tilde W^{\epsi,n}_{\rm sc} = W(-i\epsi\nabla_k)$,
where $\nabla_k$ is understood with periodic boundary conditions on $\R^d/\Gamma^*$.
\begin{Lem}\label{difference_eps}
For any $n\in\mathcal{I}$
\begin{equation}\label{Wdiag}
\tilde W^{\epsi,n} = \tilde W^{\epsi,n}_{\rm sc} + o(\epsi)
\end{equation}
in $B(L^2(M^*))$.
\end{Lem}
\begin{proof} 
By definition we have 
\[
\left( \tilde W^{\epsi,n}_{\rm sc} \psi \right)(k) = (2\pi)^{-d/2}
\int_{\mathbb{R}^d} \widehat W^\epsi (p)\psi_n(k-p) \,dp\,,
\]
and therefore
\begin{eqnarray}\label{wdiff} 
\lefteqn{\left(
\left( \tilde{W}^{\epsi,n} - 
\tilde{W}^{\epsi,n}_{\rm sc} \right)
\psi\right) (k) =} \nonumber\\
& = 
(2\pi)^{-d/2} {\displaystyle \int_{\mathbb{R}^d}} \widehat W^{\epsi,n}(p)
\Big( \big(\varphi_n(k),\varphi_n(k-p)\big)_{L^2(M)} -1\Big)\psi(k-p)\, dp\,.
\end{eqnarray}
As to be shown, there exists a constant $c$ such that
\begin{equation}\label{skalar}
\left|\big(\varphi_n(k),\varphi_n(k-p)\big)_{L^2(M)} -1\right| \leq c|p|^2
\end{equation}
for Lebesgue almost all $k$.
Therefore we  conclude
\begin{eqnarray*}
\left\| \left(\tilde{W}^{\epsi,n} - 
\tilde{W}^{\epsi,n}_{\rm sc}\right)
 \psi \right\|_{L^2(M^*)} & \leq &
c\epsi^2\left\| \int \left| \widehat W^\epsi (p) \frac{|p|^2}{\epsi^2}\right|
|\psi(k-p)|\, dp\,\right\|_{L^2(M^*)} \\
&\leq& 
c' \epsi^2 \|\psi\|_{L^2(M^*)}\,.
\end{eqnarray*}

To show \eqref{skalar} note that one can chose $\varphi_n(k)$ such that 
the map $k\mapsto \varphi_n(k)\in L^2(M)$ is smooth Lebesgue almost everywhere.
This is because according to Lemma \ref{smooth_projection} the projections
$P_n(k)$ depend smoothly on $k$ and hence one can locally define
$\varphi_n(k) = P_n(k) \varphi_n(k_0)/\| P_n(k) \varphi_n(k_0) \|$.
Now we can cover $M^*$ by finitely many open disjoint sets $U_i$ such that 
$M^*\setminus \cup_i U_i$ is a set of Lebesgue measure zero and $\varphi_n(k)$ can be defined 
on the closure of each $U_i$ in the way described above. 
One obtains
a family $\varphi_n(k)$ of eigenfunctions which is smooth except at the 
boundaries between the sets, where we pick $\varphi_n(k)$ with an arbitrary phase.
Wherever $\varphi_n(k)$ is smooth,
Taylor expansion yields $\varphi_n(k-p) = \varphi_n(k) -p \cdot \nabla_k
\varphi_n(k) + \frac{1}{2} p\cdot\mathbb{H}(\varphi_n)(k')p$, 
where $\mathbb{H}(\varphi_n)$ denotes the Hessian
and $\frac{1}{2} p\cdot \mathbb{H}(\varphi_n)(k')p$ is the Lagrangian remainder.
In view of  $(\varphi_n(k),\nabla_k\varphi_n(k))_{L^2(M)}=0$, which follows 
from comparing  (\ref{gradient_projection}) with 
\[
(\nabla_k P_n \psi)(k) = ( \varphi_n(k),\psi (\cdot,k)) \nabla_k \varphi_n(k) +
( \nabla_k \varphi_n(k),\psi(\cdot,k))\varphi_n(k)\,,
\]
 we obtain 
\begin{equation*}
\left|\big(\varphi_n(k),\varphi_n(k-p)\big)_{L^2(M)} -1\right| \leq c(k)|p|^2\,.
\end{equation*}
Here $c(k) = \frac{1}{2} \sum_{i,j} |(\varphi_n(k'),\partial_{k_i}\partial_{k_j}
\varphi_n(k'))|$. 
However, $c(k)$ is bounded uniformly in $k$, since $\varphi_n(k)$ is smooth on each
compact $\bar {U_i}$. 
\end{proof}

We  define now  the semiclassical Hamiltonian
$H^{\epsi,n}_{\rm sc}$   
\begin{equation}
H^{\epsi,n}_{\rm sc} = E_n(k) + W(-i\epsi\nabla_k)
\end{equation}
acting on $L^2(M^*)$. 
Then Lemma \ref{difference_eps} shows that the difference 
$H^{\epsi,n} -  H^{\epsi,n}_{\rm sc}$ is of order $o(\epsi)$
uniformly in $B(L^2(M^*))$ and hence (cf.\ Section \ref{effective}) 
the difference of the corresponding unitary groups
approaches zero as $\epsi\to 0$.
\begin{Cor}\label{ucor}
Let $U^{\epsi,n}_\mathrm{sc}(t)=
\mathrm{e}^{-it{H}^{\epsi,n}_\mathrm{sc}}$ and
${U}^{\epsi,n} (t)=
\mathrm{e}^{-it{H}^{\epsi,n}}$, then
\[
\lim_{\epsi\to 0}\big({U}^{\epsi,n}(t/\epsi)-
{U}^{\epsi,n}_\mathrm{sc}(t/\epsi)\big)=0
\]
in $B(L^2(M^*))$.
\end{Cor} 

The semiclassical limit for $U^{\epsi,n}_{\rm sc}(t/\epsi)$
on $L^2(\mathbb{T}^*)$ is well studied. We refer to \cite{folland,hoevermann,robert}. As a 
consequence the strong limits
\begin{eqnarray}\label{62}
\lim_{\epsi\to 0} U^{\epsi,n}_{\rm sc}(- t/\epsi)\,(-i\epsi \nabla_k)\,  
U^{\epsi,n}_{\rm sc}(t/\epsi) & = & r_n(t;k)\,, \\ \label{61}
\lim_{\epsi\to 0}  U^{\epsi,n}_{\rm sc}(- t/\epsi)\,k\,  
U^{\epsi,n}_{\rm sc}(t/\epsi) & = & k_n(t;k)
\end{eqnarray}
exist on $H^1(\mathbb{T}^*)$. $r_n$ and $k_n$ act as multiplication operators and are defined as
in (\ref{dynamical_system}) with initial conditions $(r_n(0),k_n(0))=(0,k)$. 

Since the restriction of $\epsi x^{n}_{\rm diag}$ to the $n$-th band subspace
is unitarily equivalent to $- i\epsi \nabla_k$ on $L^2(\mathbb{T}^*)$, 
we can, in view of Theorem \ref{MT2}, conclude the proof of Theorem \ref{MT3} by
showing
\begin{Lem} In $B(L^2(M^*))$ we have  
\begin{equation} \label{xx1}
\lim_{\epsi\to 0}
\big( U^{\epsi,n}(- t/\epsi)(-i\epsi \nabla_k)  
U^{\epsi,n}(t/\epsi) - 
 U^{\epsi,n}_{\rm sc}(- t/\epsi)(-i\epsi \nabla_k)  
U^{\epsi,n}_{\rm sc}(t/\epsi)
\big) =0\,.
\end{equation}
\end{Lem}
\begin{proof}
The proof of (\ref{xx1}) 
is  analogous to the proof of Theorem \ref{MT2} in Section 5, however, simpler.
As in (\ref{mean_momentum}) we have
\[
 U_{\rm sc}^{\epsi,n}(- t/\epsi)(-i\epsi \nabla_k)  
U_{\rm sc}^{\epsi,n}(t/\epsi) = -i\epsi\nabla_k
+ \epsi \int_0^{t/\epsi}  U_{\rm sc}^{\epsi,n}(-s) 
\big[ -i\nabla_k, H_{\rm sc}^{\epsi,n} \big]
  U_{\rm sc}^{\epsi,n}(s)\,ds
\]
and 
\begin{eqnarray*}\lefteqn{
 U^{\epsi,n}(- t/\epsi)(-i\epsi \nabla_k)  
U^{\epsi,n}(t/\epsi) =}\\
& = & -i\epsi\nabla_k 
+ \epsi \int_0^{t/\epsi}  U^{\epsi,n}(-s) 
\big[ -i\nabla_k, H^{\epsi,n} \big]
  U^{\epsi,n}(s)\,ds\\
& = &  -i\epsi\nabla_k
+ \epsi \int_0^{t/\epsi}  U^{\epsi,n}(-s) 
\Big( \big[ -i\nabla_k, H^{\epsi,n}_{\rm sc} \big]+
\big[ -i\nabla_k, \Delta \tilde W^{\epsi,n}\big]
\Big) U^{\epsi,n}(s)\,ds\,,
\end{eqnarray*}
where $ \Delta \tilde W^{\epsi,n} :=\tilde{W}^{\epsi,n} - 
\tilde{W}^{\epsi,n}_{\rm sc}$. Now $[ -i\nabla_k, H^{\epsi,n}_{\rm sc}] 
= -i\nabla_k E_n(k)$ is bounded, and
(\ref{xx1}) follows from Corollary \ref{ucor} if we can show that
$[ -i\nabla_k, \Delta \tilde W^{\epsi,n}] = o(1)$ in $B(L^2(M^*))$.
Noting that $ (\Delta \tilde W^{\epsi,n}\psi)(k)$ is given by
(\ref{wdiff}), this can be shown by an argument similar to the one in Lemma
\ref{difference_eps}.

\end{proof}

\section{Semiclassical equations of motion for general observables} \label{MT4S}

We proceed to more general semiclassical observables. First note that
Theorem \ref{MT5} follows immediately from the results obtained so far
(Theorem \ref{MT1}, Corollary \ref{ucor} and \eqref{61}), since
multiplication with $k$ in Bloch representation is bounded.
Hence we now have that
\begin{equation}
\lim_{\epsi \to 0} \|x^\epsi(t)\psi - \mathcal{U}^{-1}R(t)\mathcal{U}\psi\|=0
\end{equation}
for all $\psi \in \mbox{Ran} P_\mathcal{I}\cap D(|x|)\cap H^2$ and that
\begin{equation}
\lim_{\epsi \to 0} \|k^\epsi(t)\psi - \mathcal{U}^{-1} K(t)\mathcal{U}\psi\| = 0
\end{equation}
for all $\psi \in \mbox{Ran}  P_\mathcal{I}$.
We next consider bounded continuous functions of $x^\epsi(t)$ and $k^\epsi(t)$:
\begin{Lem} \label{fl}
Let $f\in C_\infty(\R^d)$ and $g\in C(M^*)$. Then 
for all $\psi\in\mbox{Ran}P_\mathcal{I}$ we have 
\begin{equation} \label{fxcon}
\lim_{\epsi\to 0}\|\left( f(x^\epsi(t)) - \mathcal{U}^{-1}f(R(t))\mathcal{U}\right) \psi\|=0
\end{equation}
and
\begin{equation} \label{fkcon}
\lim_{\epsi\to 0}\|\left( g(k^\epsi(t)) - \mathcal{U}^{-1}g(K(t))\mathcal{U}\right) \psi\|=0\,.
\end{equation}
\end{Lem}
\begin{proof}
We will sketch the proof for $x^\epsi(t)$. First note that 
$\bar R(t):=\mathcal{U}^{-1}R(t)\mathcal{U}$
is a bounded self-adjoint operator and commutes with $P_\mathcal{I}$.
Hence the sets $D_\pm := (\bar R(t)\pm i)(\mbox{Ran} P_\mathcal{I}\cap D(|x|)\cap H^2)$
are dense in $P_\mathcal{I}$ (Since $R$ and $x^\epsi$ are vectors of operators in 
$\R^d$, note that this and the following statements hold component wise).
For $\psi \in D_\pm$ we have
\begin{equation} \label{Rescon}
\left[ (x^\epsi(t) \pm i)^{-1} - (\bar R(t) \pm i)^{-1}\right]\psi =
 (x^\epsi(t) \pm i)^{-1} (\bar R(t) - x^\epsi(t))\varphi
\end{equation}
for $\varphi = (\bar R(t) \pm i)^{-1} \psi \in \mbox{Ran} P_\mathcal{I}\cap D(|x|)\cap H^2$.
Thus, by Theorem \ref{MT1}, \eqref{Rescon} strongly approaches zero as $\epsi\to 0$ and,
since $D_\pm$ are dense in $P_\mathcal{I}$,   $(x^\epsi(t) \pm i)^{-1}$ strongly approach
$(\bar R(t) \pm i)^{-1}$ on $P_\mathcal{I}$.

Using the fact that polynomials in $(x_j\pm i)^{-1}$, $j=1,\ldots,d$, are dense in $C_\infty(\R^d)$ 
one concludes
that the convergence
$x^\epsi(t)\to\bar R(t)$ on Ran$P_\mathcal {I}$ in the ``strong resolvent sense'' implies
\[
\lim_{\epsi\to 0} \|\left(f(x^\epsi(t))-f(\bar R(t))\right)\psi\| =0
\]
for all $f\in C_\infty(\R^d)$ and $\psi \in \mbox{Ran}P_\mathcal{I}$ 
(cf.\ Theorem VIII.20 in \cite{reed-simon-i}).
However, by the functional calculus for self-adjoint operators we have
$f(\mathcal{U}^{-1}R(t)\mathcal{U}) = \mathcal{U}^{-1}f(R(t))\mathcal{U}$ and 
\eqref{fxcon} follows.

Clearly \eqref{fkcon} follows analogously.
\end{proof}
\begin{proof}[Proof of Theorem \ref{MT4}]
Let $a\in \mathcal{O}(0)$. Referring again to the general Stone-Weier\-stra\ss\ theorem we
can uniformly approximate $a(x,\xi)$ by a sum of products, i.e.\ 
$a(x,\xi)= \sum_{i=0}^\infty a_i f_i(x)g_i(\xi)$ with $f_i\in C_\infty(\R^d)$, 
$g_i\in C(M^*)$, $\sum |a_i|<\infty$ and
$\sup_{i\in \mathbb{N}, x\in \R^d, \xi \in M^*}$ $|f_i(x)g_i(\xi)|<\infty$. Hence in order
to prove Theorem \ref{MT4} we are left to show that for arbitrary $f\in C_\infty(\R^d)$
and $g\in C(M^*)$ we have
\begin{equation} \label{fe}
\left(f(x)g(\xi)\right)^{W,\epsi}(t) \to \mathcal{U}^{-1}f(R(t)) g(K(t))\mathcal{U}
\end{equation}
strongly on Ran$P_\mathcal{I}$.  
To see this recall the so called product rule for quantum observables (cf.\ \cite{robert}).
It states, in particular, that for two symbols $A,B\in \mathcal{O}(0)$ 
\[
\lim_{\epsi\to 0} \left\|\left( (AB)^{W,\epsi} - A^{W,\epsi}B^{W,\epsi}\right)\psi\right\| = 0\,.
\] 
Applied to our case this yields
\[
\left(f(x)g(\xi)\right)^{W,\epsi}(t) \to \left(f(x)^{W,\epsi}g(\xi)^{W,\epsi}\right)(t)
= f(x^\epsi(t))g(k^\epsi(t))\,.
\]
Finally, since $f$ and $g$ are bounded, Lemma \ref{fl} implies \eqref{fe} and thus 
Theorem \ref{MT4}.
\end{proof}

\section{Band crossings}\label{outlook}

We proved the semiclassical limit for isolated bands only. In principle, there are two 
distinct mechanisms of how this assumption could be violated. First of all a band could 
be isolated but have a constant multiplicity larger than one. This occurs, e.g., for the 
Dirac equation where because of spin the electron and positron bands are both two-fold 
degenerate. A systematic study is only recent \cite{gerard-markowich-mauser-poupaud,spohn2} 
and leads to a matrix valued 
symplectic structure for the semiclassical dynamics. For periodic potentials degeneracies 
are the exception. 
They form a real analytic subvariety 
of the Bloch variety $B=\{(k,\lambda)\in\mathbb{R}^d\times\mathbb{R}\mid\exists f\in L^2(M): 
H_{\mathrm{per}}(k)f=\lambda f\}$
and have a dimension at least one 
less than the dimension of $B$ \cite{kuchment,wilcox}.
Thus points of band crossings have a $k$-Lebesgue measure 
zero. From the study of band structures in solids one knows that band crossings indeed occur. 
Thus it is of interest 
to understand the extra complications coming from band crossings.

There are two types of band crossings. The first one is removable through a proper analytic 
continuation of the bands. In a way, removable band crossings correspond to a wrong choice 
of the fundamental domain. E.g.\ for $V=0$ we may artificially introduce a lattice $\Gamma$. 
The bands touch then at the boundary of $M^*$. Upon analytic continuation we recover the 
single band $E_1(k)=k^2/2$ with $M^*=\mathbb{R}^d$. In one dimension all band crossings can 
be removed \cite{reed-simon-iv}. Thus, with the adjustment discussed, our result fully covers 
the case $d=1$. For $d\geq 2$ generically band crossings cannot be removed.

 It is then of great physical interest to understand how a wave packet tunnels 
into a neighboring band through points of degeneracy (or almost degeneracy). For a careful 
asymptotic analysis in particular model systems we refer to the monumental work of 
G.\ Hagedorn \cite{hagedorn}. Gerard \cite{gerard} considers a model system with two bands in 
two dimensions, i.e., the role of $-\frac{1}{2}\Delta+V$ is taken by 
$\left(\begin{array}{cc} k_1 & k_2 \\ k_2 & -k_1 \end{array}\right)$. He investigates the 
semiclassical limit and proves that the particle may tunnel to the other band with a probability
which depends on how well the initial wave packet is concentrated near a semiclassical orbit hitting
the singularity. 

\section*{Acknowledgments}
FH gratefully acknowledges the financial support by the Deutsche Forschungsgemeinschaft via 
the Graduiertenkolleg {\em Mathematik im Bereich ihrer Wechselwirkung mit der Physik} at the 
LMU M\"{u}nchen.


\begin{thebibliography}{99}
\bibitem{asch-knauf} J.\ Asch and A.\ Knauf. {\em Motion in Periodic Potentials}, 
Nonlinearity {\bf 11}, 175-200 (1998).
\bibitem{ashcroft-mermin} N.\ W.\ Ashcroft and N.\ D.\ Mermin. {\em Solid State Physics}, 
Saunders (1976).
\bibitem{avron-elgart} J.\ E.\ Avron and A. Elgart. {\em Adiabatic Theorem without a Gap 
Condition}, Commun.\ Math.\ Phys.\ {\bf 203}, 445-463 (1999).
\bibitem{avron-seiler-yaffe} J.\ E.\ Avron, R.\ Seiler and L.\ G.\ Yaffe. {\em Adiabatic 
Theorems and Applications to the Quantum Hall Effect}, Commun.\ Math.\ Phys.\ {\bf 110}, 
33-49 (1987).
\bibitem{buslaev} V.\ Buslaev. {\em Semiclassical Approximation for Equations with Periodic 
Coefficients}, Russ.\ Math.\ Surveys {\bf 42}, No.\ 6, 97-125 (1987).
\bibitem{buslaev-grigis} V.\ Buslaev and A.\ Grigis. {\em Imaginary Parts of Stark-Wannier 
Resonances}, J.\ Math.\ Phys.\ {\bf 39}, No. 5, 2520-2550 (1998).
\bibitem{cycon-froese-kirsch-simon} H.\ L.\ Cycon, R.\ G.\ Froese, W.\ Kirsch and B.\ 
Simon. {\em Schr\"{o}dinger Operators}, Springer (1987).
\bibitem{folland} G.\ B.\ Folland. {\em Harmonic Analysis in Phase Space}, Princeton 
University Press (1989).
\bibitem{gerard-markowich-mauser-poupaud} P.\ Gerard, P.\ A.\ Markowich, N.\ J.\ Mauser and 
F.\ Poupaud. {\em Homogenization Limits and Wigner Transforms}, Commun.\ Pure Appl.\ Math. 
{\bf 50}, 323-380 (1997).
\bibitem{gerard-martinez-sjostrand} C.\ Gerard, A.\ Martinez and J.\ Sj\"{o}strand. 
{\em A Mathematical Approach to the Effective Hamiltonian in Perturbed Periodic Problems}, 
Commun.\ Math.\ Phys. {\bf 142}, 217-244 (1991).
\bibitem{gerard}  P.\ Gerard. {\em Semiclassical Limits}, talk at {\em Nonlinear Equations in
Many-Particle Systems}, Oberwolfach (1999).
\bibitem{guillot-ralston-trubowitz} J.\ C.\ Guillot, J.\ Ralston and E.\ Trubowitz. 
{\em Semi-Classical Asymptotics in Solid State Physics}, Commun.\ Math.\ Phys.\ {\bf 116}, 
401-415 (1988).
\bibitem{hagedorn} G.\ A.\ Hagedorn. {\em Molecular Propagation through Electron Energy 
Level Crossings}, Memoirs Am.\ Math.\ Soc.\ {\bf 111} (1994).
\bibitem{hoevermann} F.\ H\"{o}vermann. {\em Quantum Motion in Periodic Potentials}, 
Dissertation, LMU M\"{u}nchen (1999)
\bibitem{kato} T.\ Kato. {\em Perturbation Theory for Linear Operators}, Springer (1980).
\bibitem{kohn} W.\ Kohn. {\em Theory of Bloch Electrons in a Magnetic Field: The Effective 
Hamiltonian}, Phys.\ Rev.\ {\bf 115}, No.\ 6, 1460-1478 (1959).
\bibitem{kuchment} P.\ Kuchment. {\em Floquet Theory for Partial Differential Equations}, 
Birkh\"{a}user (1993).
\bibitem{markowich-mauser-poupaud} P.\ A.\ Markowich, N.\ J.\ Mauser and F.\ Poupaud. 
{\em A Wigner-function Theoretic Approach to (Semi)-Classical Limits: Electrons in a 
Periodic Potential}, J.\ Math.\ Phys. {\bf 35}, No.\ 3, 1066-1094 (1994).
\bibitem{nenciu} G.\ Nenciu. {\em Dynamics of Band Electrons in Electric and Magnetic 
Fields: Rigorous Justification of the Effective Hamiltonians}, Rev.\ Mod.\ Phys.\ {\bf 63}, 
No.\ 1, 91-127 (1991).

\bibitem{reed-simon-i} M.\ Reed and B.\ Simon. {\em Methods of Modern Mathematical 
Physics I}, Academic Press (1972).
\bibitem{reed-simon-iv} M.\ Reed and B.\ Simon. {\em Methods of Modern Mathematical 
Physics IV}, Academic Press (1978).
\bibitem{robert} D.\ Robert. {\em Autour de l'Approximation Semi-Classique}, Birkh\"{a}user 
(1987).
\bibitem{spohn} H.\ Spohn. {\em Long Time Asymptotics for Quantum Particles in a Periodic 
Potential}, Phys.\ Rev.\ Lett.\ {\bf 77}, No.\ 7, 1198-1201 (1996).
\bibitem{spohn2} H.\ Spohn. {\em Semiclassical Limit of the Dirac Equation and Spin 
Precession}, Annals of Physics, to appear.
\bibitem{wilcox} C.\ H.\ Wilcox. {\em Theory of Bloch Waves}, J.\ Anal.\ Math. {\bf 33}, 
146-167 (1978).
\bibitem{zak} J.\ Zak. {\em Dynamics of Electrons in Solids in External Fields}, Phys.\ 
Rev.\ {\bf 168}, No.\ 3, 686-695 (1968).
\end{thebibliography}
\end{document}